\documentclass[runningheads]{llncs}
\usepackage[T1]{fontenc}

\usepackage{graphicx}
\usepackage{tabularx}
\usepackage{booktabs}
\usepackage{multirow}
\usepackage{booktabs}
\usepackage{siunitx}
\usepackage{amsmath,amssymb,amsfonts}
\usepackage[table,xcdraw]{xcolor}
\usepackage[colorlinks]{hyperref}
\usepackage{color, colortbl}
\usepackage{footmisc}
\usepackage{subfig}
\usepackage{float}

\begin{document}

\title{The Effect of Lossy Compression on 3D Medical Images Segmentation with Deep Learning}
\titlerunning{Effect of Lossy Compression on 3D Medical Images Segmentation}



\author{
    Anvar Kurmukov \inst{4} \and 
    Bogdan Zavolovich \inst{1} \and
    Aleksandra Dalechina \inst{1} \and
    Vladislav Proskurov \inst{4} \and
    Boris Shirokikh \inst{2, 3}
}

\authorrunning{A. Kurmukov et. al}

\institute{
    National Research Nuclear University MEPhI
    \and
    Skolkovo Institute of Science and Technology, Moscow, Russia
    \and
    Artificial Intelligence Research Institute (AIRI), Moscow, Russia
    \and
    Independent researcher
    \\
}

\maketitle
\begin{abstract}
Image compression is a critical tool in decreasing the cost of storage and improving the speed of transmission over the internet. While deep learning applications for natural images widely adopts the usage of lossy compression techniques, it is not widespread for 3D medical images.  Using three CT datasets (17 tasks) and one MRI dataset (3 tasks) we demonstrate that lossy compression up to $\times20$ times have no negative impact on segmentation quality with deep neural networks (DNN). In addition, we demonstrate the ability of DNN models trained on compressed data to predict on uncompressed data and vice versa with no quality deterioration.
\end{abstract}


\section{Introduction}
The advent of digital technology has revolutionized medical imaging, leading to an exponential increase in the volume of clinical data generated and stored globally. As the digital archives of medical images continue to expand, managing the storage requirements of this quantity of data has become a pressing challenge. Conventional approaches to addressing this problem have included the use of lossless compression techniques, such as JPEG and JPEG2000, which have been employed in an attempt to mitigate the issue of disk space scarcity \cite{foos2000jpeg}.

A more memory-efficient approach is to use lossy compression regimes; however, their application is not without consequences. Particularly in the context of deep learning (DL), where data management and loading efficiency are critical to model performance and GPU utilization, authors have demonstrated models' performance deterioration when trained on heavily compressed images ~\cite{gandor2022first,10184656}.

Recent studies, predominantly in non-medical domains, have investigated the impact of lossy image compression on the performance of neural networks, revealing a nuanced relationship between compression and model efficacy. Research by \cite{benbarrad2021impact,junior2023assessing} on object detection with convolutional neural networks (CNNs) and subsequent inquiries into applications ranging from steel surface detection to agricultural image segmentation have illustrated both the potential advantages and drawbacks of incorporating lossy compression into data preprocessing workflows. These discoveries emphasize the importance of balancing data efficiency with the preservation of model accuracy and reliability.

Despite the growing body of research in various domains, the exploration of the impact of image compression on medical imaging applications, particularly in the realm of deep learning-based analysis, remains relatively uncharted. Previous studies have primarily focused on the direct effects of compression on image quality rather than its implications for algorithmic performance \cite{liu2017current}. However, emerging evidence from the analysis of histopathological images for metastatic cancer \cite{ghazvinian2019impact} detection and mammogram classification \cite{jo2021impact} suggests that, under certain conditions, models trained on compressed images may not only maintain but potentially enhance their performance on compressed data, challenging preconceived notions about the incompatibility of lossy compression and high-stakes medical image analysis.

In our work, we examine how lossy compression affects DL-based segmentation models on 20 segmentation tasks on 3D computed tomography (CT) and magnetic resonance imaging (MRI) data. Our key contributions are three-fold:
\begin{enumerate}
\item Through systematic experiments on four medical image datasets, we demonstrate that storage memory footprint could be reduced at training time up to 20 times via lossy compression with no negative effect on segmentation quality.
\item We illustrate that DNN models based on Unet architecture trained on uncompressed data do not deteriorate while inferenced on compressed data, and vice versa.
\item We show that while trained experts could accurately differentiate between lossy compressed images after a compression rate of 10 times, DL models still preserve their quality for images compressed 20 times.
\end{enumerate}

\section{Methods}
\subsection{JPEG2000}
The Joint Photographic Experts Group (JPEG) developed the JPEG 2000 image compression standard, which employs discrete wavelet transformation. This approach ensures the preservation of high image quality during the compression process. A key feature of JPEG 2000 is its support for progressive loading, enabling the stepwise display of image details as more data is received. Furthermore, JPEG 2000 provides the flexibility to either maintain all image details with lossless compression or manage the level of lossy compression according to specific needs. This functionality has been incorporated into the DICOM standard as Supplement 61, substantially enhancing the efficiency of medical image processing and storage. Despite high compression ratio, JPEG 2000 retains critical image details when required, demonstrating an optimal balance between storage efficiency and image quality \cite{foos2000jpeg}.

\subsection{Data}
We used four publicly available datasets, covering two of the most popular 3D medical imaging modalities: computed tomography (CT) and magnetic resonance imaging (MRI), and two types of segmentation tasks: anatomical (15 organs and one anatomical structure) and pathological region segmentation (brain tumors and liver tumors).

\subsubsection{LiTS}
We utilized 131 training CT images with liver tumor segmentation masks from the LiTS dataset \cite{lits}. We cropped all images to a given liver mask; the rest of the preprocessing was done in nnunet.

\subsubsection{BraTS}
We employed 1251 training skull-stripped MRI cases from the BraTS21 dataset \cite{brats}. Each training case includes native (T1) and post-contrast T1-weighted (T1Gd) images, T2-weighted and T2 Fluid Attenuated Inversion Recovery (T2-FLAIR) volumes. The BraTS21 dataset contains annotations of the necrotic tumor core, the peritumoral invaded tissue, and GD-enhancing tumor.

\subsubsection{AMOS}
We utilized 300 CT images from the training part of the AMOS dataset \cite{amos}. We performed a multi-class segmentation task, predicting 15 abdominal organ masks: spleen, right kidney, left kidney, gallbladder, esophagus, liver, stomach, aorta, inferior vena cava, pancreas, right adrenal gland, left adrenal gland, duodenum, bladder, and prostate/uterus.

\subsubsection{BGPD}
We employed the Burdenko Glioblastoma Progression Dataset (BGPD) \cite{BGPD}, a systematically curated collection of patients diagnosed with primary glioblastoma. For our experiments, we used 176 CT topometric images, predicting segmentation masks of the brain stem.

\subsection{Experiments}

We selected nnU-Net \cite{nnunet} for our experiments.
It is an open-source framework\footnote{\url{https://github.com/MIC-DKFZ/nnUNet}}, widely adopted as a strong benchmark for medical image segmentation tasks. Low-resolution (lowres) architecture was used for primary results, fullres (default config) for ablation study. In addition, for ablation study on AMOS dataset, we used a Swin Unetr (base) \footnote{\url{https://huggingface.co/darragh/swinunetr-btcv-base}} model pretrained on BTCV dataset \cite{btcv}.

Lowres architecture has been trained for 300 epochs with an initial learning rate of $10^{-2}$ reduced to $10^{-4}$ at the last epoch. Each epoch consists of 250 iterations. Stochastic gradient descent with Nesterov momentum (0.99) is used for training. The loss function is a combination of Dice loss and cross-entropy. Poly learning rate scheduler changes the learning rate during training $LR_{new} = LR_{initial}(1 - epoch/epoch_{max})^{0.9}$. The training of 3 folds takes about 7.5 hours on a 32G NVIDIA Tesla V100.

\subsection{Metrics}

To assess the quality of the segmentation model, we used the Dice coefficient:

\begin{equation}
\text{Dice}(X, Y) = \frac{2 \times |X \cap Y|}{|X| + |Y|}
\end{equation}

where $X$ is the predicted binary segmentation, and $Y$ is the ground truth. The Dice coefficient for multi-class segmentation is then obtained by averaging the Dice scores across all segmentation classes.


We used the Peak Signal-to-Noise Ratio (PSNR) to assess the quality of the compressed images. A higher value indicates a closer match between the original and compressed images. Given the original image $I_1$ and the compressed image $I_2$ of size $M \times N$, PSNR is calculated based on the mean squared error (MSE):

\begin{equation}
\text{MSE}(I_1, I_2) = \frac{\sum_{M, N} (I_1(m, n) - I_2(m, n))^2}{M \times N}
\end{equation}

The PSNR is then defined as:

\begin{equation}
    \text{PSNR}(I_1, I_2) = 10 \cdot \log_{10} \left(\frac{r^2}{\text{MSE}(I_1, I_2)}\right),
\end{equation}

where $r$ is the maximum fluctuation in the input image data type. We set $r$ to 4096 for CT data and to $\max(I) - \min(I)$ for MRI data (measuring the minimum and maximum values over the entire BraTS dataset).
\section{Results}
First, we assess how lossy compression influences images quality using human evaluation and PSNR. Second, we test the effect on segmentation quality of training a Unet architecture on lossy compressed images. Third, we test how models trained on uncompressed images performs on lossy compressed images. Finally, we demonstrate that in a scenario were storage capacity is limited, compression is a better choice then data subsampling.

\begin{figure}[ht]
\centering
\includegraphics[width=\textwidth]{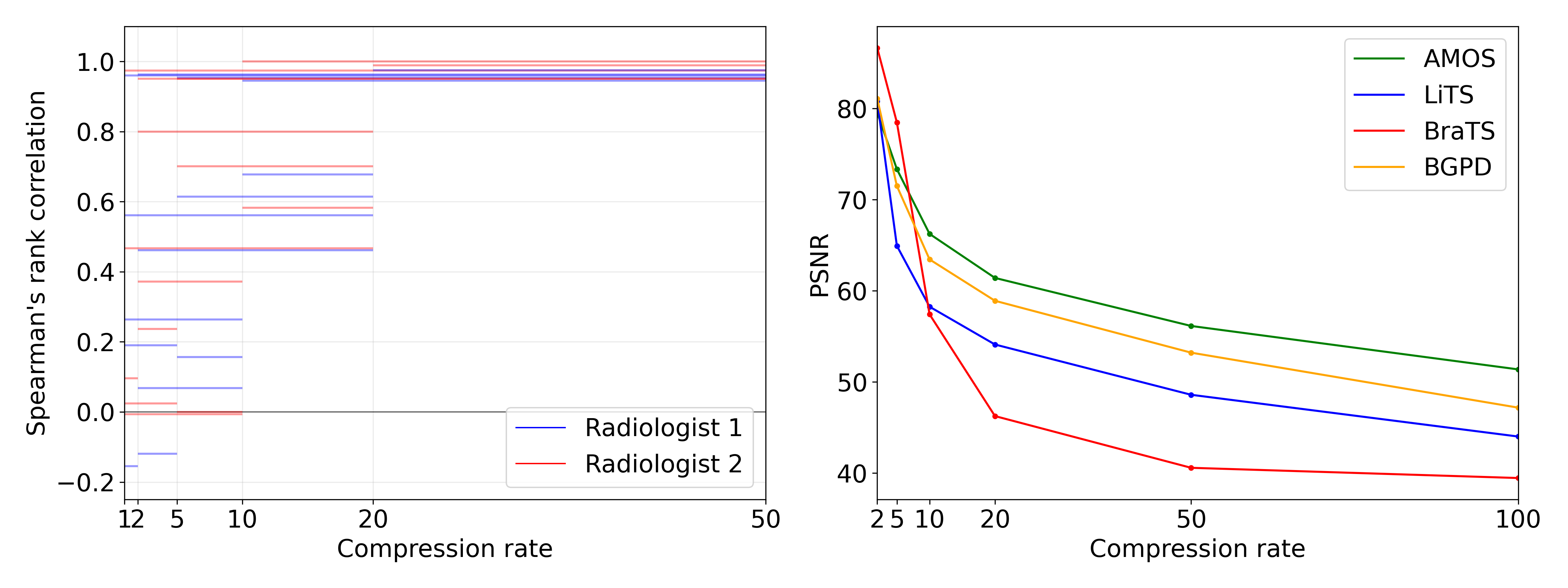}
\caption{Left: results of human assessment of image pairs with different compression ratios. Each line represents the average Spearman's rank correlation (x-axis) between two compression rates (two ends of each line, y-axis). Right: peak signal-to-noise ratio of images compressed at different rates.}
\label{fig3:human-eval-psnr}
\end{figure}

\subsection{Compressed images quality}
To assess how well radiologists differentiate between images of varying compression rates we generated pairs of images with different compression rates: 1 vs 2, 1 vs 5, ..., 20 vs 50. For each pair of compression rates (15 in total), we generated 10 random images for each dataset (600 in total). 2D axial slices with lower compression rates were randomly placed on the left or right side of the image. Each axial slice in a pair was generated from the same uncompressed 2D slice. Two radiologists (5 and 8 years of experience) independently assessed image pairs and provided a decision on which image (left or right) was of better quality or if they could not decide which one was better. We next compare their results with  the ground truth using Spearman's Rank correlation (1 if radiologist chosen correct image of higher quality, 0 if she can not decide, -1 if image of lower quality is selected as image of higher quality). See example images in the Supplementary materials.

First, we observe that both radiologists are unable to distinguish between uncompressed images and images with compression rates up to 5. Their performance steadily improves at compression rates 10 and 20, achieving almost perfect quality for compression rate 50 Figure ~\ref{fig3:human-eval-psnr} (left). Figure ~\ref{fig3:human-eval-psnr} (right) shows the PSNRs achieved via JPEG2000 lossy compression at different compression rates. We observe a significant drop in PSNR at rate 10, which then steadily declines. This result intuitively aligns with the outcomes from human evaluation.

\subsection{Effect of training on lossy compressed images}
\label{sec:trainin-compressed}
\begin{figure}[ht!]
\centering
\includegraphics[width=\textwidth]{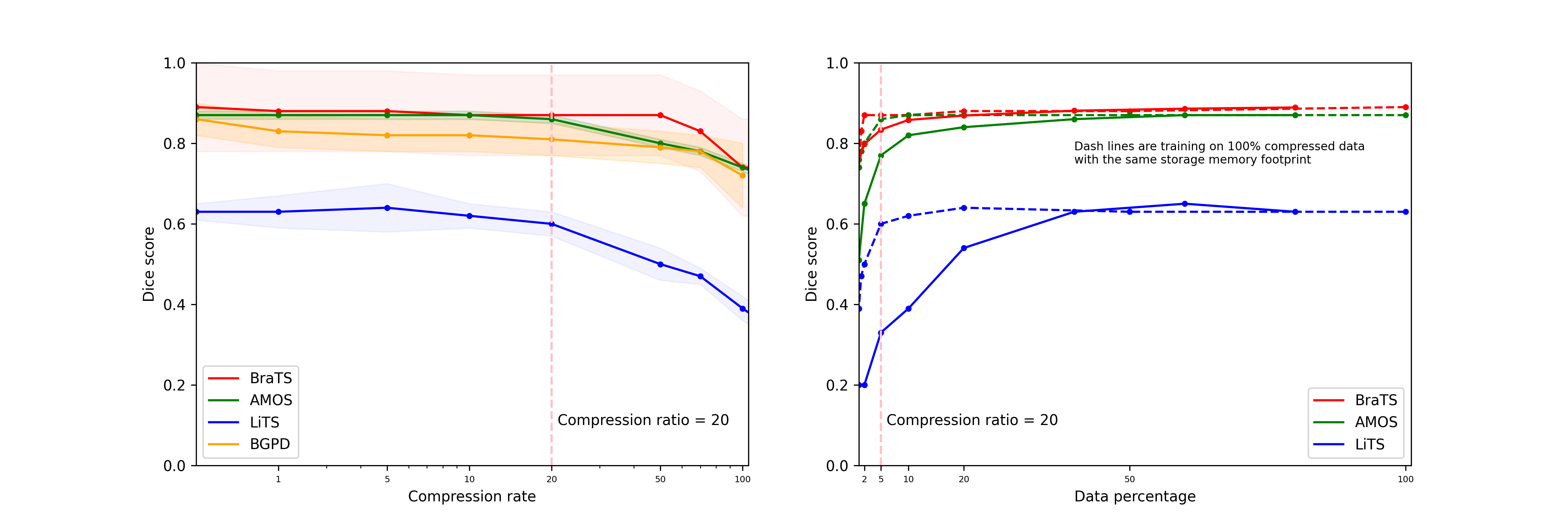}
\caption{Left: segmentation performance of models trained on data compressed at different rates. Right: storage memory footprint tradeoff, few-shot (uncompressed) vs full-shot (compressed).}
\label{fig2:few-shot-seg-per}
\end{figure}

To investigate the influence of image compression on the downstream segmentation task, we trained the Unet segmentation architecture on images compressed at different rates and on uncompressed original images (separate models) Figure~\ref{fig2:few-shot-seg-per} (left). For all 4 datasets, we observe stable segmentation quality up to compression rates 20, with rapid deterioration of the quality at higher compression rates. This finding provides strong evidence that the Unet architecture with a standard training pipeline is robust to artifacts introduced by the JPEG2000 compression algorithm. Table~\ref{tab:amos-by-class} presents the by-class segmentation performance on the AMOS dataset. We observe no segmentation drop, even for smaller organs like adrenal glands and the gallbladder.

\begin{table}
\centering
\caption{Per-organ segmentation quality for different compression rates on AMOS. Numbers are mean Dice score averaged over 3-fold cross-validation.}
\begin{tabular}{l@{\hspace{15pt}}c@{\hspace{15pt}}c@{\hspace{15pt}}c@{\hspace{15pt}}c@{\hspace{15pt}}c@{\hspace{15pt}}c@{\hspace{15pt}}c@{\hspace{15pt}}c}
    \toprule
    compression rate        & 1    & 2    & 5    & 10   & 20   & 50   & 70   & 100\\
    \midrule
    SPL     & 0.96 & 0.96 & 0.96 & 0.96 & 0.96 & 0.94 & 0.92 & 0.90\\
    RKI     & 0.84 & 0.84 & 0.84 & 0.84 & 0.83 & 0.76 & 0.72 & 0.67\\
    LKI     & 0.76 & 0.76 & 0.76 & 0.76 & 0.73 & 0.63 & 0.59 & 0.54\\
    GBL     & 0.76 & 0.76 & 0.76 & 0.76 & 0.75 & 0.62 & 0.57 & 0.51\\
    ESO     & 0.80 & 0.80 & 0.80 & 0.80 & 0.78 & 0.70 & 0.67 & 0.62\\
    LIV     & 0.87 & 0.87 & 0.88 & 0.87 & 0.85 & 0.79 & 0.76 & 0.73\\
    STO     & 0.82 & 0.81 & 0.81 & 0.82 & 0.80 & 0.75 & 0.73 & 0.70\\
    AOR     & 0.96 & 0.96 & 0.96 & 0.96 & 0.95 & 0.93 & 0.91 & 0.89\\
    IVC     & 0.95 & 0.95 & 0.95 & 0.95 & 0.95 & 0.92 & 0.91 & 0.89\\
    PAN     & 0.82 & 0.81 & 0.82 & 0.82 & 0.79 & 0.66 & 0.61 & 0.52\\
    RAD     & 0.84 & 0.84 & 0.84 & 0.84 & 0.83 & 0.78 & 0.75 & 0.72\\
    LAD     & 0.97 & 0.97 & 0.97 & 0.97 & 0.97 & 0.96 & 0.95 & 0.94\\
    DUO     & 0.90 & 0.90 & 0.90 & 0.90 & 0.89 & 0.85 & 0.83 & 0.80\\
    BLA     & 0.95 & 0.95 & 0.95 & 0.95 & 0.95 & 0.93 & 0.92 & 0.90\\
    PRO/UTE & 0.90 & 0.90 & 0.90 & 0.90 & 0.89 & 0.85 & 0.83 & 0.80\\
    \midrule
    average     & 0.87 & 0.87 & 0.87 & 0.87 & 0.86 & 0.80 & 0.78 & 0.74\\
    \bottomrule
\end{tabular}
\label{tab:amos-by-class}
\end{table}
\subsection{Inference models trained on original data}
Next, we examine how models trained on images of higher compression rates perform on images of lower compression rates, as shown in Figure~\ref{fig4:inference-amos-lits}. On the AMOS and LiTS datasets, we observe no performance drop for compression rates up to 20. The selection of compression rate 20 is due to the fact that beyond this rate, models' performance decreases compared to models trained on uncompressed data, as discussed in Section~\ref{sec:trainin-compressed}. This scenario is important if one wants to use models already trained on compressed data on uncompressed data, potentially saving time on compression itself.

Furthermore, we investigated how models trained on uncompressed data perform on data with various compression rates. For this experiment, we utilized the Unet (trained via the full resolution nn-Unet pipeline) and publicly available weights for the SWIN Unetr model (trained on the BTCV dataset
The results are presented in Table~\ref{tab:amos-infer-2-5-20}. Once again, we observe no deterioration in segmentation quality for either model. The significance of this result lies in the fact that researchers already have a plethora of trained models (both public and private), typically on uncompressed data. Our findings suggest that adopting an "inference on compressed data strategy" does not introduce any additional costs. Despite relatively modest (by an absolute value) segmentation performance of pretrained SWIN Unetr\footnote{Likely due to the fact it was trained on a BTCV dataset, using only 24 training samples.}, the purpose of this experiment was to demonstrate that JPEG compression has almost no effect on the performance of publicly available network trained on an uncompressed data.

\begin{table}
\centering
\caption{Models trained on uncompressed data tested on compressed data. Results on AMOS dataset. nn-Unet results are averaged over three cross validation folds. Swin Unetr results are averaged over all AMOS dataset.}
\begin{tabular}{l@{\hspace{15pt}}c@{\hspace{15pt}}c@{\hspace{15pt}}c@{\hspace{15pt}}c@{\hspace{15pt}}c}
    \toprule
    compression rate        & 2    & 5    & 10   & 20    & 50 \\
    \midrule
    nnUnet (high resolution)     & 0.971 & 0.971 & 0.970 & 0.969 & 0.946\\
   Swin Unetr \cite{hatamizadeh2022swin}     & 0.614 & 0.614 & 0.614 & 0.614& 0.603 \\
    \bottomrule
\end{tabular}
\label{tab:amos-infer-2-5-20}
\end{table}



\subsection{Few-shot vs lossy compressed full training}
Finally, we examine the memory footprint tradeoff of training on compressed images. We trained nn-Unet in a few-shot regime, using 1\%, 5\%, 10\%, 20\%, 50\%, and 100\% of the data. We then compared the segmentation quality of the resulting algorithms using the same holdout set. Figure~\ref{fig2:few-shot-seg-per} (right) illustrates the Dice score of networks trained in a few-shot regime. Dashed lines visualize the segmentation quality of the networks trained on 100\% of the data at different compression rates with the same memory footprint. For example, training a network on all training data compressed 20 times is equivalent (in terms of storage memory) to training on 5\% of uncompressed data.

The results on all 4 datasets suggest that at a compression rate of 20, training on 100\% compressed data is superior compared to training the network on a fraction of uncompressed data with the same storage memory footprint. 

\begin{figure}[ht]
\centering
\includegraphics[width=\textwidth]{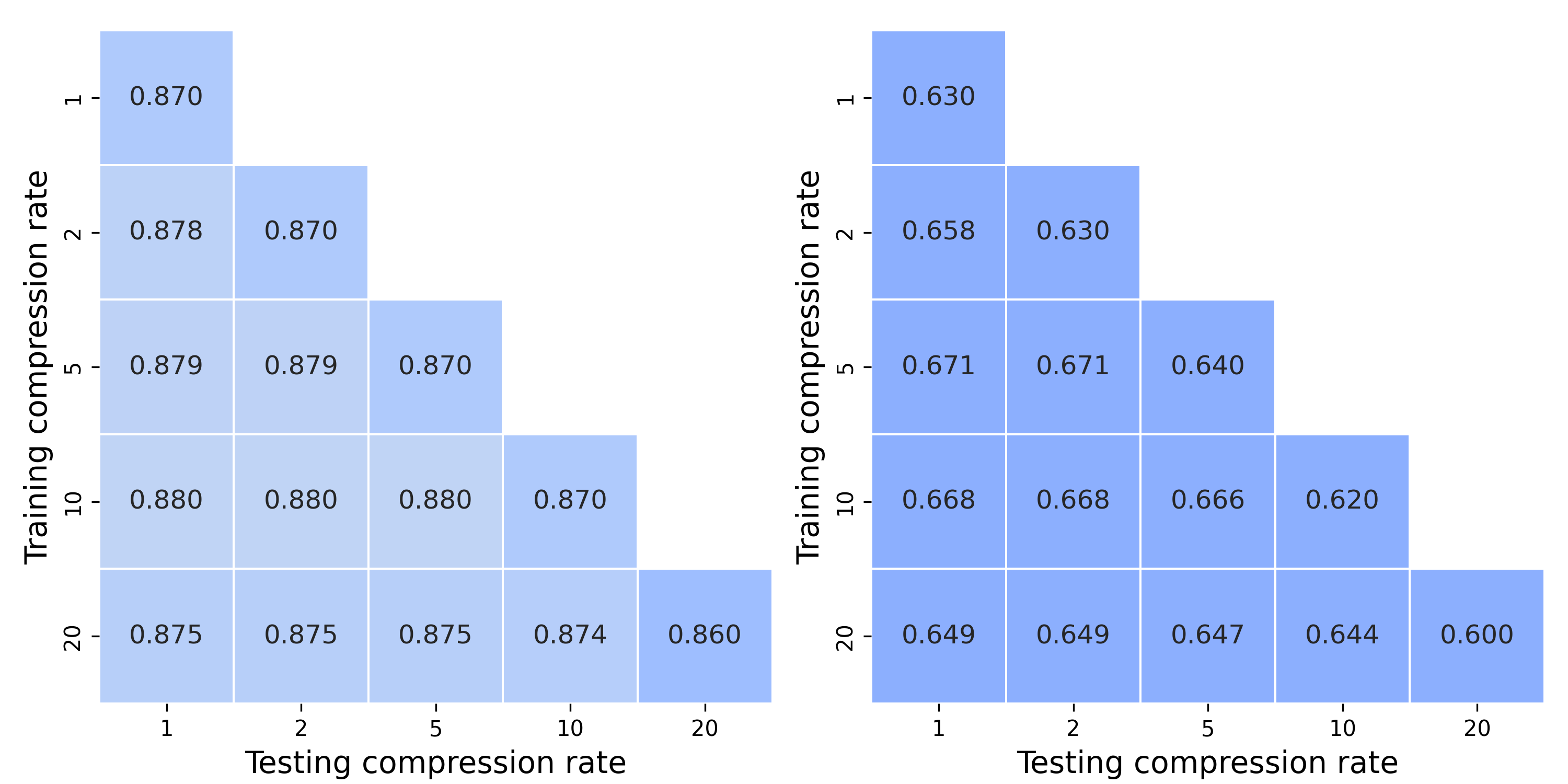}
\caption{Testing model trained on images at different compression rates on images of other comression rates. Results for AMOS dataset (left), LiTS dataset (right). }
\label{fig4:inference-amos-lits}
\end{figure}

\section{Conclusion}
In this study, we conducted a comprehensive investigation into the impact of lossy image compression on the performance of deep learning-based segmentation models in the context of 3D medical imaging. Our findings demonstrate that significant reductions in storage memory footprint can be achieved through the use of lossy compression without compromising the segmentation quality of the DL models. We hope that our work will encourage medical imaging researchers, workig with 3D images to start integrating lossy compression in their pipelines, without the unreasonable fear of (segmentation) quality loss. 

\subsection{Study limitations}
First, we solely tested the effect of lossy compression using JPEG2000. While we believe this choice is justified by the fact that this is the only lossy compression algorithm currently supported by DICOM, investigating the effect of other algorithms might be interesting. Second, we only focus our experiments on a single neural network architecture Unet, with a limited analysis of a publicly available pretrained Swin Unetr. Finally, human evaluation experiment only tested the ability of radiologists to differentiate between images of varying compression, but not their performance at segmenting images after compression.


\bibliographystyle{splncs04}
\bibliography{bibliography.bib}

\newpage
\section*{Supplementary materials}

\begin{figure}[ht]
\centering
\includegraphics[width=\textwidth]{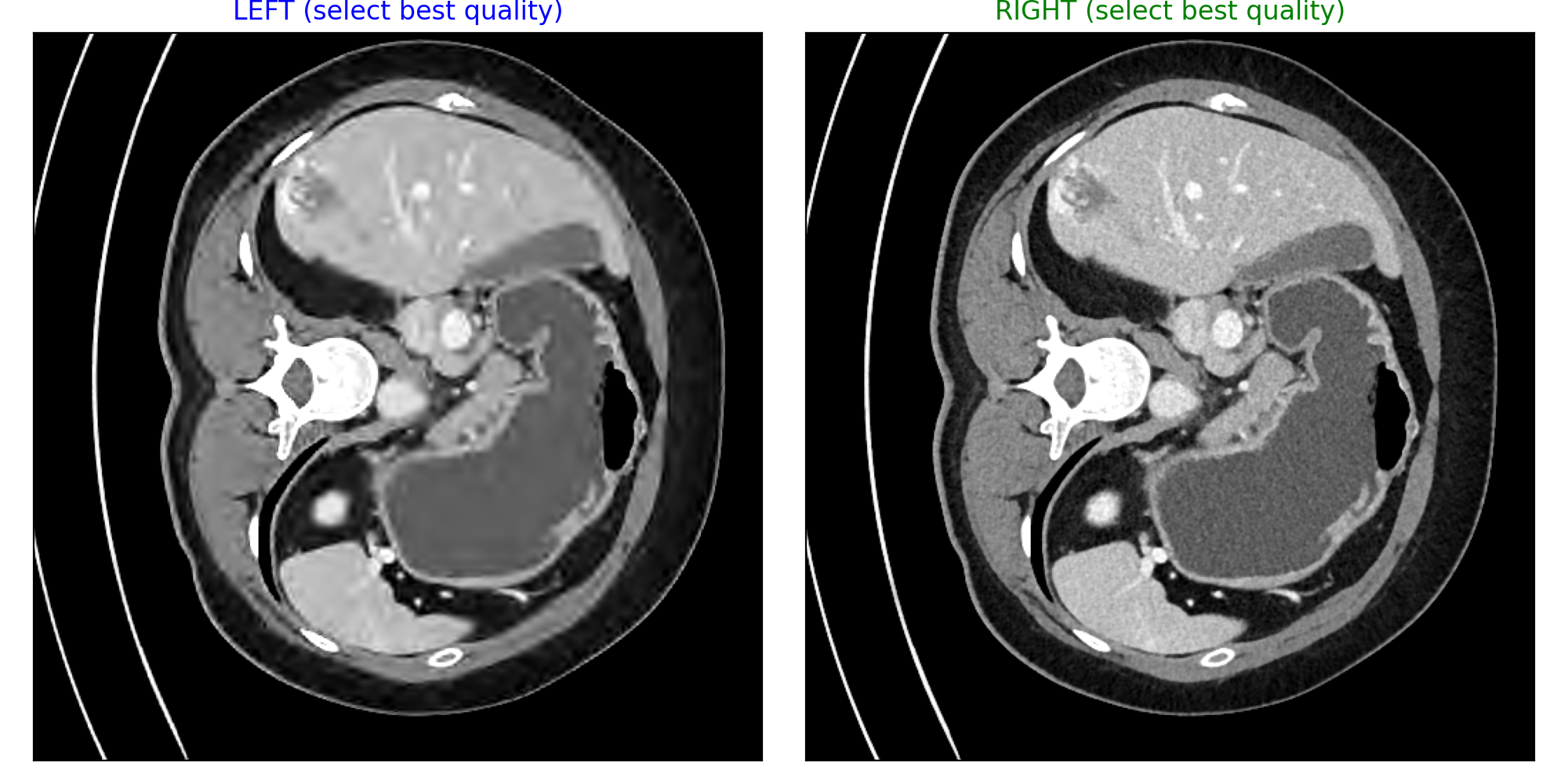}
\caption{Example image used during human evaluation. Radiologists were asked to decide whether left or right image looks of higher quality (or both looks indistinguishable). CT images were clipped to abdominal, brain, or liver window respectively.}
\label{fig4:inference-amos-lits}
\end{figure}

\begin{figure}[ht]
\centering
\includegraphics[width=\textwidth]{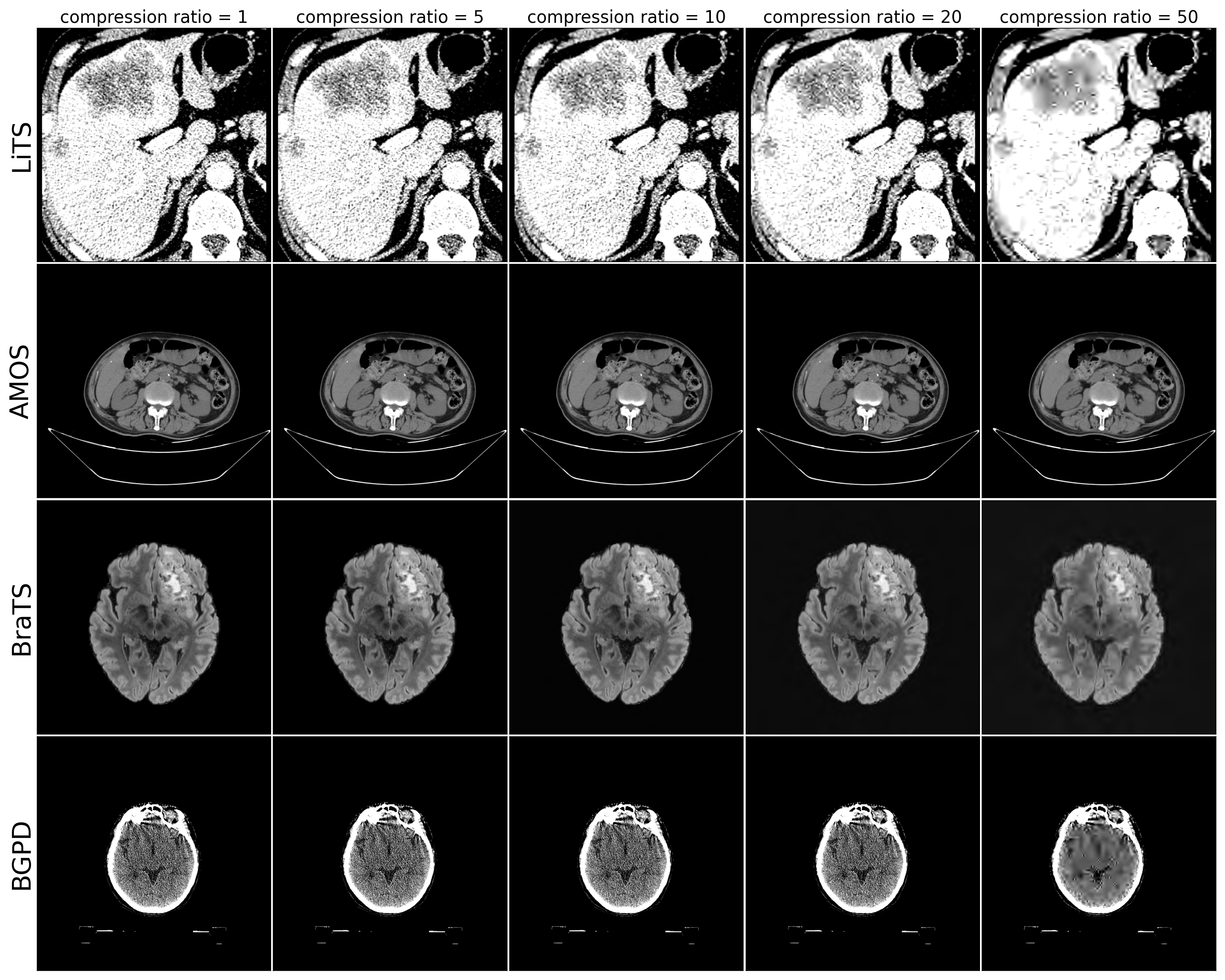}
\caption{Effect of JPEG2000 compression on images from different experimental datasets.}
\label{fig4:inference-amos-lits}
\end{figure}

\end{document}